\documentclass{revtex4}
\usepackage{amssymb,amsmath}
\usepackage{comment}
\usepackage[retainorgcmds]{IEEEtrantools}
\usepackage{graphicx}
\usepackage{epstopdf}
\usepackage{float}
\usepackage{listings}
\usepackage{color}
\usepackage{xcolor}

\usepackage[pdftex,                
bookmarks=true,
bookmarksnumbered=true,
bookmarksopen=true, %
bookmarksopenlevel=\maxdimen,%
hypertexnames=false,
breaklinks=true,
pdfpagemode=UseOutlines,
linkbordercolor={1 0 0}]{hyperref} 

\usepackage{chngcntr}

\begin{document}

\title{The Non-Adiabatic Pressure Perturbation and Non-Canonical Kinetic Terms in Multifield Inflation}

\author{Carsten van de Bruck}
\email{C.vandeBruck@sheffield.ac.uk}
\affiliation{Consortium for Fundamental Physics, School of Mathematics and Statistics, \\
		University of Sheffield, Hounsfield Road, Sheffield, S3 7RH, United Kingdom}

\author{Susan Vu}
\email{susan.vu@sheffield.ac.uk}
\affiliation{Consortium for Fundamental Physics, School of Mathematics and Statistics, \\
		University of Sheffield, Hounsfield Road, Sheffield, S3 7RH, United Kingdom}

\date{\today}

\begin{abstract}
The evolution of the non-adiabatic pressure perturbation during inflation driven by two scalar fields is studied numerically for three different types of 
models. In the first model, the fields have standard kinetic terms. 
The other two models considered feature non-canonical kinetic terms; the first containing two fields which are coupled via their kinetic terms, and the second where one field has the standard kinetic term with the other field being a DBI field.
We find that the evolution and the final amplitude of the non-adiabatic pressure perturbation 
depends strongly on the kinetic terms.
\end{abstract}

\maketitle

\section{Introduction}
Inflation is the most successful theory when describing primordial perturbations in the universe.
These primordial density perturbations are generated from quantum fluctuations of the field which has driven inflation (or, as 
an extension, of the perturbations in the curvaton field). This theory is consistent with observations. 
Details about the nature of the primordial perturbations depend on the details of the theory, such as 
whether there was only one field responsible for inflation, the form of the potential and interactions between fields. 
Observations of the cosmic microwave background radiation (CMB) 
constrain the amount of isocurvature perturbations to high accuracy (of order of 10\% of the curvature perturbations) and thereby ruling out a large number of inflationary models. 

It has been recently pointed out that one important quantity to be studied in detail is the non--adiabatic pressure perturbation. Apart 
from being an essential ingredient to determine the evolution of the curvature perturbation, it was shown to source vorticity perturbations 
at second order in perturbation theory \cite{ChristMalik09}. To be more precise, the evolution of the second order vorticity $\omega_{2ij}$ was shown to obey

\begin{equation}
\omega_{2ij}' - 3{\cal H} c_s^2 \omega_{2ij}= \frac{2a}{\rho+P}\left[3{\cal H} V_{1[i}\delta P_{{\rm nad}1,j]} +  \frac{\delta\rho_{1,[j}\delta P_{{\rm nad},i]}}{\rho + P} \right]
\end{equation}
where on the right hand side the first order quantities $\delta \rho_1$ (the energy density perturbation at first order) and $\delta P_{\rm nad}$ (the non-adiabatic 
pressure perturbation) appear. If the non-adiabatic pressure perturbation vanishes at first order, the vorticity decays not only at first order, but also at second order. However, 
if $\delta P_{\rm nad}$ is non-zero, which is usually the case in multifield inflation, vorticity at second order is sourced by the non-adiabatic pressure perturbation. 
This opens up more possibilities of tests of inflation, as the non-zero vorticity can source $B$-mode polarisation of the CMB photons. 
It is therefore important to understand the behaviour of $\delta P_{\rm nad}$ in diverse models. In addition to sourcing vorticity, the non--adiabatic perturbation affects also the evolution of the curvature perturbation on super--horizon scales, and thereby influencing the predictions for non--Gaussianity in these models, see e.g. \cite{Cai:2009hw,Emery:2012sm} and references therein.

In \cite{Hust12}, the evolution of the non-adiabatic pressure perturbation has been studied in detail for theories with canonical kinetic terms with different choices of potentials. 
In this paper, we consider the evolution of $\delta P_{\rm nad}$ in models with non-canonical kinetic terms. In particular we consider a theory in which the kinetic terms are coupled 
as in \cite{Branden03, Wands96, Choi07}, and a model in which one field has the standard form with the other being a DBI field (see e.g. \cite{SilvTong04,AlishaSilv04,Peiris:2007gz} and references therein).

The paper is organised as follows: In the next Section (\ref{sec:models}) we write down the theories considered in this paper and derive the evolution equations for the perturbations. 
In Section \ref{sec:results} we present the results of our numerical calculations. Our conclusions are presented in Section \ref{sec:conclusion}.

\section{The models}
\label{sec:models}
We will consider three models in this paper. The actions are given as follows:
\begin{enumerate}
	\item Two of the models we consider are described by actions of the form 
		\begin{equation}
			\mathcal{S} =  \int d^4 x\sqrt{-g} \bigg[ \frac{M_{Pl}^2}{2}R - \frac{1}{2} g^{\mu \nu} \partial_\mu \phi \partial_\nu \phi - \frac{1}{2} e^{2b(\phi)} g^{\mu \nu} \partial_\mu 						\chi \partial_\nu \chi - V(\phi,\chi) \bigg]~,
		\end{equation}
where in the first model $b(\phi)=0$ and in the second model 
\begin{equation}
b(\phi) = \beta \phi \nonumber
\end{equation}
and $ \beta $ is a constant. 

\item The third model we consider contains a scalar field with a canonical kinetic term and one DBI field. The action is given by
\begin{equation}
\mathcal{S} =  \int d^4 x\sqrt{-g} \bigg[ \frac{M_{Pl}^2}{2}R - \frac{1}{2} g^{\mu \nu} \partial_\mu \phi \partial_\nu \phi - \frac{1}{f(\chi)} ( 1 - \gamma^{-1} ) - V(\phi,\chi) \bigg]~,
\end{equation}
where
\begin{equation}
\gamma = \frac{1}{ \sqrt{1 + f(\chi) g^{\mu \nu} \partial_\mu\chi \partial_\nu\chi  } } , \qquad f(\chi) = \frac{\lambda}{ (\chi^2 + \mu^2)^2 }. \nonumber
\end{equation}
$\gamma$ is the warp factor describing the shape of the extra dimensions. Both $ \lambda $ and $ \mu $ are constants.
\end{enumerate}
In all the actions given, $ M_{Pl} = (8 \pi G)^{-1/2}$ is the reduced Planck mass and $R$ is the Ricci scalar.

\subsection{Background Equations of Motion}
We assume a spatially flat Friedmann-Robertson-Walker, FRW, spacetime
	\begin{equation}
		ds^2 = -dt^2 + a^2(t) \delta_{ij} dx^i dx^j
	\end{equation}
where $ a(t) $ is the scale factor.

\subsubsection{Kinetic Coupling}
The equations of motion for both fields and the Friedmann equations are given by \cite{LalakLang07,Branden03}
	\begin{align}
		\ddot{\phi} + 3H \dot{\phi} + V_{\phi} = {} & b_{\phi} e^{2b} \dot{\chi}^2 \\
		\ddot{\chi} + ( 3H + 2b_{\phi} \dot{\phi} ) \dot{\chi} + e^{-2b} V_{\chi} = {} &  0
	\end{align}
and
	\begin{align}
		3 H^2 = {} & \frac{1}{2} ( \dot{\phi}^2 + e^{2b} \dot{\chi}^2 ) + V \\
		-2 \dot{H} = {} & \dot{\phi}^2 + e^{2b} \dot{\chi}^2
	\end{align}
where $V_{\phi} = dV/d\phi$ and similarly for $V_{\chi}$, $ b_{\phi} = db(\phi)/d\phi $ and the Hubble parameter is given as $ H = \dot{a}/a $.
These equations reduce to the two standard field case when the kinetic coupling is set to zero.

\subsubsection{DBI Field}
The equations of motion for the fields for this model are
	\begin{align}
		\ddot{\phi} + 3H \dot{\phi} {} & + V_{\phi} =  0 \\
		\ddot{\chi} + 3H \gamma^{-2} \dot{\chi} {} & + \frac{1}{2} \frac{f_{\chi}}{f^2} ( 1 - 3 \gamma^{-2} + 2 \gamma^{-3} ) + \gamma^{-3} V_{\chi} =  0 
	\end{align}
and Einstein's equations give
	\begin{align}
		3 H^2 = {} & \frac{1}{2} \dot{\phi}^2 + \frac{1}{f}( \gamma - 1 ) + V \\
		-2 \dot{H} = {} & \dot{\phi}^2 + \gamma \dot{\chi}^2
	\end{align}
where $ f_{\chi} = df/d\chi $. The dots represent differentiation with respect to cosmic time $ t $.

\subsection{First Order Perturbation Equations}
We turn now our attention to the first order perturbation equations and work in the longitudinal gauge, in which the line element is given by 
\begin{equation}
ds^2 = -(1-2\Psi)dt^2 + a^2(1+2\Psi) \delta_{ij}dx^i dx^j~.
\end{equation}
First, we decompose the fields into their homogeneous and perturbed parts
\begin{equation}
\phi(t,\bold{x}) = \phi(t) + \delta \phi (t, \bold{x}), \qquad \chi(t,\bold{x}) = \chi(t) + \delta \chi (t, \bold{x}) 
\end{equation}
and we will be working with the Fourier components of the perturbations, $ \delta \phi_{\bold{k}} (t) $ and $ \delta \chi_{\bold{k}} (t) $ and will 
be omitting the subscript $ \bold{k} $ to shorten the expressions.
\newline

The perturbed Einstein field equations for the model concerning the scalar field and the DBI field are
\begin{align}
\dot{\Psi} + H \Psi = {} & \frac{1}{2} ( \dot{\phi} \delta \phi + \gamma \dot{\chi} \delta \chi ) , \\
\ddot{\Psi} + 4H\dot{\Psi} + ( 2\dot{H} + 3H^2 )\Psi = {} &\frac{1}{2} \bigg[ \dot{\phi} \delta \dot{\phi} - \dot{\phi}^2 \Psi - V_{\phi} \delta \phi - V_{\chi} \delta \chi - \frac{1}{2} \frac{f_{\chi}}				{f^2} \bigg( 2 - \frac{1}{\gamma} - \gamma \bigg) \delta \chi + \gamma( \dot{\chi} \delta \dot{\chi} - \dot{\chi}^2 \Psi ) \bigg] , \\
3H( H\Psi + \dot{\Psi} ) + \frac{k^2}{a^2} \Psi = {} & - \frac{1}{2} \bigg[ \dot{\phi} \delta \dot{\phi} - \dot{\phi}^2 \Psi + V_{\phi} \delta \phi + V_{\chi} \delta \chi +  \frac{1}{2} \frac{f_{\chi}}				{f^2} \bigg( 2 - 3\gamma + \gamma^3 \bigg) \delta \chi - \gamma^3( \dot{\chi} \delta \dot{\chi} + \dot{\chi}^2 \Psi ) \bigg] .
\end{align}

We find the Klein-Gordon equations for the field perturbations
\begin{align}
\delta \ddot{\phi} {} & + 3H \delta \dot{\phi} + \bigg( \frac{k^2}{a^2} + V_{\phi \phi} \bigg) \delta \phi + V_{\phi \chi} \delta \chi - 4 \dot{\phi} \dot{\Psi} + 2 V_{\phi} \Psi = 0 \\
\delta \ddot{\chi} {} & +  3\bigg( H + \frac{\dot{\gamma}}{\gamma} \bigg) \delta \dot{\chi} +  \bigg[ \frac{k^2}{a^2 \gamma^2} + \frac{V_{\chi \chi}}{\gamma^3} + \frac{f_{\chi} \dot				{\gamma}}{f \gamma} \dot{\chi} - \frac{1}{2} \frac{f_{\chi}}{\gamma} \dot{\chi}^2 V_{\chi} 
		+ \frac{1}{2} \bigg(1 - \frac{1}{\gamma} \bigg)^2 \bigg( \frac{1}{\gamma} \bigg( \frac{f_{\chi}}{f^2} \bigg)_{,\chi} + \bigg(1 + \frac{1}{\gamma} 		\bigg)\frac{1}{f} \bigg( \frac{f_{\chi}}{f} \bigg)_{,\chi} \bigg) \bigg] \delta \chi \nonumber \\
	& + \frac{V_{\chi \phi}}{\gamma^3} \delta \phi - \bigg( \frac{3}{\gamma^2} + 1 \bigg) \dot{\chi}\dot{\Psi} + \bigg[ \frac{f_{\chi}}{f^2 \gamma^3} (1 - \gamma )^2 + \frac{V_{\chi}}			{\gamma^3}(1 + \gamma^2) - 2\frac{\dot{\gamma}}{\gamma}\dot{\chi} \bigg] \Psi = 0
\end{align}

It is convenient to work with the gauge-invariant Mukhanov-Sasaki variables \cite{Sas86,Muk88}, defined by 
\begin{equation}
Q_{\phi} \equiv  \delta \phi + \frac{\dot{\phi}}{H} \Psi , \qquad Q_{\chi} \equiv  \delta \chi + \frac{\dot{\chi}}{H} \Psi~.
\end{equation}

The gauge-invariant form of the perturbation equations are
\begin{align}
\label{perturbationdbi1}
\ddot{Q}_{\phi} {} & + 3H\dot{Q}_{\phi} +B_{\phi}\dot{Q}_{\chi} + \bigg(\frac{k^{2}}{a^2} + C_{\phi \phi} \bigg)Q_{\phi} + C_{\phi \chi}Q_{\chi} = 0 , \\
\label{perturbationdbi2}
\ddot{Q}_{\chi} {} & + \bigg{(}3H + 3\frac{\dot{\gamma}}{\gamma} \bigg{)} \dot{Q}_{\chi} + B_{\chi}\dot{Q}_{\phi} + \bigg{(}\frac{k^{2}}{a^2 \gamma^{2}} + C_{\chi \chi}\bigg{)}Q_{\chi} + C_{\chi \phi}Q_{\phi} = 0,
\end{align}
with the coefficients $B_{\phi}, B_{\chi}, C_{\phi \phi}, C_{\phi \chi}, C_{\chi \chi}, C_{\chi \phi} $, in the equations are as follows
\begin{align}
B_{\phi} = {} & \frac{\dot{\phi}}{2H}\gamma^{3}\bigg{(}1-\frac{1}{\gamma^{2}}\bigg{)}\dot{\chi} , \\
B_{\chi} = {} &-\frac{\dot{\phi}}{2H}\bigg{(}1 - \frac{1}{\gamma^{2}}\bigg{)}\dot{\chi} , \\
C_{\phi \phi} = {} & 3\dot{\phi}^{2} - \gamma^{3}\bigg{(}1 + \frac{1}{\gamma^{2}} \bigg{)}\frac{\dot{\phi}^{2}\dot{\chi}^{2}}{4H^{2}} - \frac{\dot{\phi}^{4}}{2H^{2}} + 2\frac{\dot{\phi}}{H}V_				{,\phi} + V_{\phi \phi} , \\
C_{\phi \chi} = {} &\frac{\dot{\phi}}{4H}\frac{f_{\chi}}{f^{2}}\frac{1}{\gamma}(1-\gamma)^{2}(\gamma^{2} + 2\gamma -1) +3\gamma\dot{\chi}\dot{\phi} - \frac{\gamma^{4}}{4H^{2}}\bigg				{(}1+\frac{1}{\gamma^{2}}\bigg{)}\dot{\phi}\dot{\chi}^{3} - \frac{\gamma\dot{\phi}^{3}\dot{\chi}}{2H^{2}} + \frac{\dot{\phi}}{H}V_{,\chi} + \frac{\gamma\dot{\chi}}{H}		V_{\phi} + V_{\phi \chi} , \\
C_{\chi \chi} = {} &\frac{1}{H}\frac{f_{\chi}}{f^{2}}\bigg{(}1 - \frac{1}{\gamma}\bigg{)}^{2}\dot{\chi} + \bigg{(}\frac{f_{\chi}}{f} - \frac{\gamma\dot{\chi}}{H}\bigg{)}\frac{\dot{\gamma}}					{\gamma}\dot{\chi} -\frac{1}{2}\frac{f_{\chi}}{\gamma}\dot{\chi}^{2}V_{\chi} + \frac{1}{2}\bigg{(}1 - \frac{1}{\gamma}\bigg{)}^{2}\bigg{[}\frac{1}{\gamma}\bigg{(}\frac				{f_{\chi}}{f^{2}}\bigg{)}_{,\chi} + \bigg{(}1+\frac{1}{\gamma}\bigg{)}\frac{1}{f}\bigg{(}\frac{f_{\chi}}{f} \bigg{)}_{,\chi}\bigg{]} \nonumber \\
			& - \frac{\gamma^{2} \dot{\chi}^{4}}{2H^{2}} - \frac{\gamma}{4H^{2}}\bigg{(}1+ \frac{1}{\gamma^{2}}\bigg{)}\dot{\chi}^{2}\dot{\phi}^{2} + \frac{1}{H}\bigg{(}1+\frac{1}					{\gamma^{2}}\bigg{)}\dot{\chi}V_{\chi} + \frac{V_{\chi \chi}}{\gamma^{3}} + \frac{3}{2}\gamma\bigg{(}1 + \frac{1}{\gamma^{2}}\bigg{)}\dot{\chi}^{2} , \\
C_{\chi \phi} = {} &\frac{\dot{\phi}}{H}\bigg{[}\frac{1}{2}\frac{f_{\chi}}{f^{2}}\frac{1}{\gamma}\bigg{(}1 - \frac{1}{\gamma}\bigg{)}^{2} - \frac{\dot{\gamma}}{\gamma}\dot{\chi}\bigg{]} - \frac				{\gamma\dot{\phi}\dot{\chi}^{3}}{2H^{2}} + \frac{1}{2}\bigg{(}1+\frac{1}{\gamma^{2}}\bigg{)} \bigg{[} 3\dot{\phi}\dot{\chi} - \frac{\dot{\chi}\dot{\phi}^{3}}{2H^{2}} + 				\frac{\dot{\phi}}{\gamma H}V_{\chi} + \frac{\dot{\chi}}{H}V_{\phi}\bigg{]} .
\end{align}
\newline
We will later calculate the power spectrum of the curvature perturbation. The curvature perturbation is defined by 
\begin{equation}
\mathcal{R} = -\frac{H^2}{\dot{H}} \bigg( \Psi + \frac{\dot{\Psi}}{H} \bigg) + \Psi~.
\end{equation}
\indent The pressure perturbation $ \delta P $ is composed of adiabatic and non-adiabatic parts
\begin{equation}
\delta P = c_{s}^2 \delta \rho + \delta P_{nad},
\end{equation}
where $ c_s^2 = \frac{\dot{P}}{\dot{\rho}} $ is the adiabatic sound speed, $ \delta \rho $ and $ \delta P_{nad} $ are the perturbations in the energy density and non-adiabatic pressure, respectively. 

For each model the adiabatic sound speed is given as follows:
\begin{enumerate}
\item Two standard scalar fields
\begin{equation}
c_s^2 = 1 + \frac{2 ( V_{\phi} \dot{\phi} + V_{\chi} \dot{\chi} )}{ 3H ( \dot{\phi}^2 + \dot{\chi}^2 )}
\end{equation}
\item One scalar field and one with a kinetic coupling
\begin{align}
c_s^2 = {} & 1 + \frac{2 ( V_{\phi} \dot{\phi} + V_{\chi} \dot{\chi} )}{ 3H ( \dot{\phi}^2 + e^{2b} \dot{\chi}^2 )}
\end{align}
\item One scalar field and one DBI field
\begin{align}
c_s^2 = {} & \frac{2 \dot{\phi}^2 + ( 1 + \gamma^2 ) \gamma^{-1} \dot{\chi}^2}{ \dot{\phi}^2 + \gamma \dot{\chi}^2 } + \frac{2V_{\phi}\dot{\phi} + ( 1+ \gamma^2 ) \gamma^{-2} V_{\chi} \dot{\chi}}{ 3H( \dot{\phi}^2 + \gamma \dot{\chi}^2 ) } 
		 + \frac{1}{ 3H( \dot{\phi}^2 + \gamma \dot{\chi}^2 ) } \frac{ f_\chi}{f^2} \bigg( \frac{1}{\gamma} - 1\bigg)^2 \dot{\chi} - 1.
\end{align}
\end{enumerate}
For the last two models, the sound speed reduces to the two standard scalar field model when $ \beta = 0 $ and $ \gamma = 1$.
Finally, the gauge-invariant entropy perturbation \cite{Gordon00,MalikWands05} is defined as 
\begin{equation}
\mathcal{S} = \frac{H}{\dot{P}} \delta P_{nad} .
\end{equation}

Due to the complexity of these equations, for all models considered, we will evaluate them numerically following the method outlined in \cite{LalakLang07,LiddleLeach2003,vandeBruck2010,Weller:2011ey}.

\section{Results}
\label{sec:results}
We will now describe the results of our numerical calculations. To be concrete, we consider all models presented in 
Section~\ref{sec:models} with the double quadratic potential \cite{Langlois1999} which will be rewritten into the form
\begin{equation}
V(\phi,\chi) = \frac{1}{2} m_{\phi}^2 ( \phi^2 + \Gamma^2 \chi^2) 
\end{equation}
where
\begin{equation}
\Gamma = \frac{m_{\chi}}{m_{\phi}}~.
\end{equation}
As usual, $m_\phi$ and $m_\chi$ are the masses of the two scalar fields. 
All the following plots are shown in the WMAP pivot scale where $ k= 0.002 \mathrm{Mpc}^{-1} $ \cite{Komatsu2010}.

\subsubsection{Two Standard Scalar Fields}
\label{sec:twostandard}
In this model, we consider the case where $ \Gamma = 7.0 $ as studied in \cite{LalakLang07,Avgous12}. Further conditions were applied, $ m_{\phi} = 1.395 \times 10^{-6}M_{Pl} $ in order to match the numerical simulation to WMAP measurements of the curvature perturbation. Initial conditions for the two fields are $\chi_0 = \phi_0 = 12.0 $ as in \cite{Hust12}.
\begin{figure}[H]
\begin{center}
\scalebox{0.55}{\includegraphics{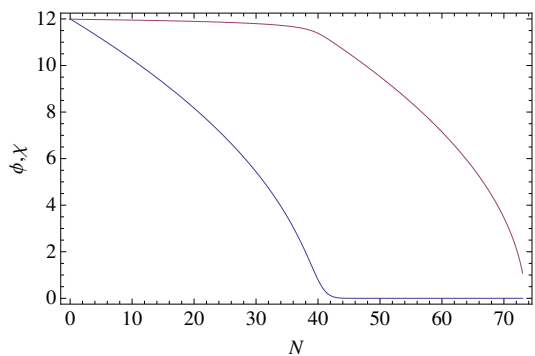}}
\caption{The background dynamics of the two fields, $\phi$-field (purple) and $\chi$-field (blue), for the two standard scalar field case.}
\label{standphichi}
\end{center}
\end{figure}
\begin{figure}[H]
\begin{center}
\scalebox{1.27}{\includegraphics{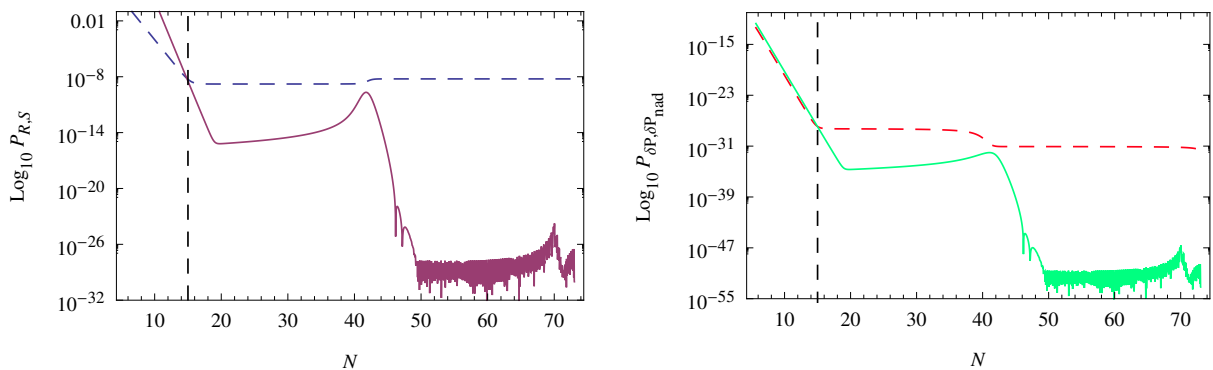}}
\caption{A comparison of the curvature and entropy power spectra, $ \mathcal{P}_{\mathcal{R}} $ (blue, dashed) and $ \mathcal{P}_{\mathcal{S}} $ (purple) on the left, and the power spectra of the total and non-adiabatic pressure perturbations, $ \mathcal{P}_{\delta P} $ (red, dashed) and $ \mathcal{P}_{\delta P_{nad}} $  (green) on the right. These are for the two standard scalar fields case. The dashed lines indicate horizon crossing.}
\label{standpow}
\end{center}
\end{figure}
We see in Figure~\ref{standphichi} that the $\chi$-field reaches the minimum of the potential, resulting in the $ \phi $-field dominating for the last 30 e-folds of inflation. This is reflected in Figure~\ref{standpow}, which at this point there is a rise in the power spectrum of the curvature perturbation. The amplitude of the entropy perturbation $ \mathcal{S} $ reduces dramatically when the exchange in field dominance occurs at 40 e-folds. It starts to gradually increase until a peak is reached at $ N = 70 $, at which a drop is experienced. At the end of inflation, we see the magnitude of the entropy perturbation is many orders smaller than the curvature perturbation, specifically $ \mathcal{S} \sim 10^{-31} $ whereas $ \mathcal{R} \sim 10^{-9} $.

The behaviour of the non-adiabatic pressure and entropy perturbation power spectra generally agree at horizon crossing at $ N = 15 $ and onwards.

We find that Figure~\ref{standpow} is in agreement with Figures 1 and 2 from \cite{Hust12}.

\subsubsection{One Standard Scalar Field and One Field containing a Kinetic Coupling}
\label{sec:kinetic}
We have two possible cases that will arise for the background; one in which the $\phi$-field reaches the minimum of the potential well before the $\chi$-field, and vice versa. We will begin by considering the latter.

In this scenario, the first 50 e-folds are dominated by the $ \phi$-field until at which point, there is an exchange in the field contributions, leaving the remaining $\chi$-field until the end of inflation.

The ratio of the field masses is $ \Gamma = 0.3 $ with $ m_{\phi} = 6.395 \times 10^{-6} $. Initial conditions are $ \chi_0 = 12.0 $ and $ \phi_0 = 11.0 $. In this case, the kinetic coupling is $ \beta = 0.1$.
\begin{figure}[H]
\begin{center}
\scalebox{0.55}{\includegraphics{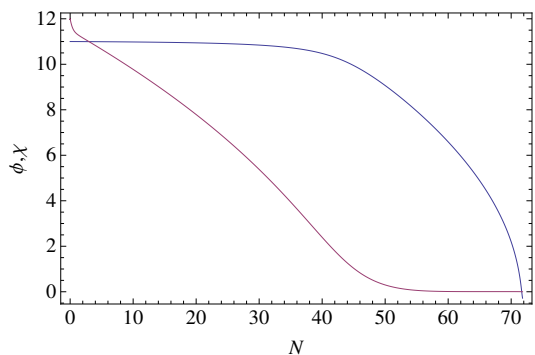}}
\caption{The background plot of the $\phi$-field (purple) and $\chi$-field (blue) for the case where $\phi$-field is initially dominant during the inflationary period. }
\label{kinetphiphichi}
\end{center}
\end{figure}

\begin{figure}[H]
\begin{center}
\scalebox{1.27}{\includegraphics{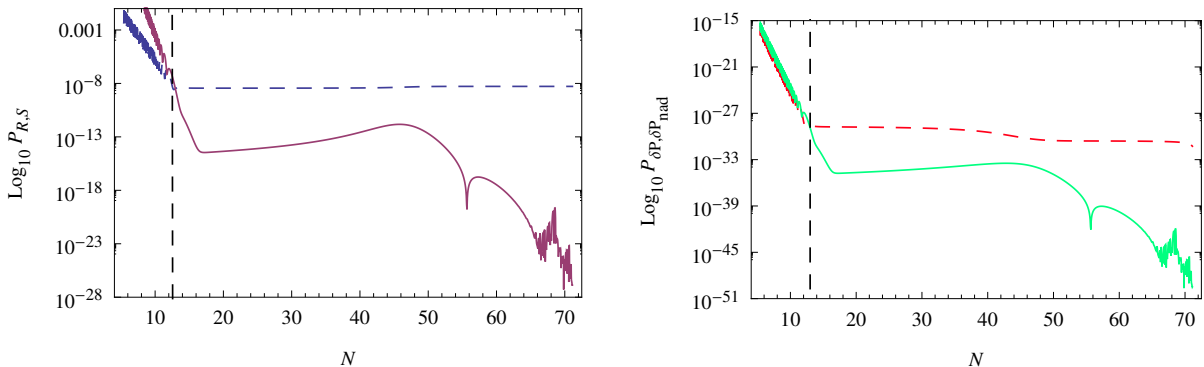}}
\caption{A comparison of various power spectra. $ \mathcal{P}_{\mathcal{R}} $ (blue, dashed) and $ \mathcal{P}_{\mathcal{S}} $ (purple) are displayed on the left and the right, $ \mathcal{P}_{\delta P} $ (red, dashed) and $ \mathcal{P}_{\delta P_{nad}} $ (green). }
\label{kinetphipow}
\end{center}
\end{figure}
The behaviour of the entropy and non-adiabatic component of the pressure perturbations is significantly different to the model previously considered in Section~\ref{sec:twostandard}. However, there is a rise and fall in $ \mathcal{S} $ during the last 6 e-folds before the end of inflation. This feature can also be seen in Figure~\ref{standpow}. 
Like the two scalar field model previously studied, the amplitude of $ \mathcal{S} $ is many times smaller than $ \mathcal{R} $. 
At the end of inflation, the final amplitudes for $ \mathcal{S} $ and $ \mathcal{P}_{\delta P_{nad}} $ are $ \mathcal{S} \sim 10^{-26} $ and $ \mathcal{P}_{\delta P_{nad}} \sim 10^{-50} $. We find that the the final entropy amplitude for this model is $ 10^{5} $ larger than found in the model only considering canonical scalar fields.

We now consider the other possible case where the $\chi$-field at first dominates the inflationary period. For this, the parameters used are $ \Gamma = 6.0 $ with $ m_{\phi} = 1.005 \times 10^{-6} $. Starting conditions for the two fields are $ \chi_0 = 7.4 $ and $\phi_0 = 7.5 $. With these parameters, the final 8 e-folds are dominated by the canonical scalar field.
\begin{figure}[H]
\begin{center}
\scalebox{0.55}{\includegraphics{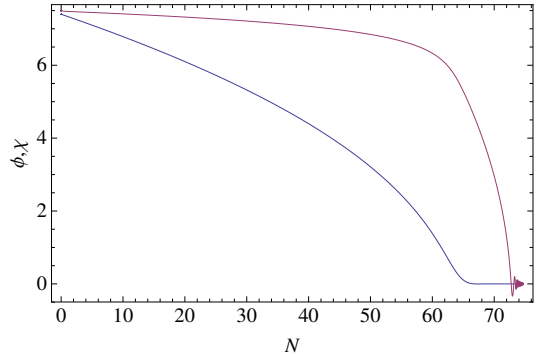}}
\caption{The background dynamics of the $\phi$-field (purple) and $\chi$-field (blue).}
\label{kinetchiphichi}
\end{center}
\end{figure}

\begin{figure}[H]
\begin{center}
\scalebox{1.27}{\includegraphics{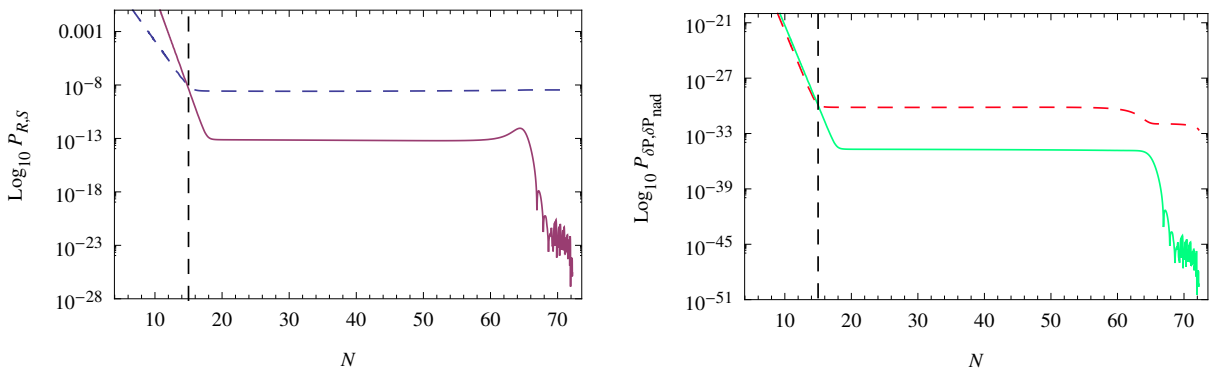}}
\caption{A comparison of the various power spectra, $ \mathcal{P}_{\mathcal{R}} $ (blue, dashed) and $ \mathcal{P}_{\mathcal{S}} $ (purple) on the left, and $ \mathcal{P}_{\delta P} $ (red, dashed) and $ \mathcal{P}_{\delta P_{nad}} $ (green) on the right. }
\label{kinetchipow}
\end{center}
\end{figure}
As expected, this choice of parameters has resulted in an one scalar field scenario at the end of inflation, similar to Section~\ref{sec:twostandard}. Due to this, the shape of the power spectra for all considered quantities will be similar. However, there is a slight difference in the last few e-folds in $ \mathcal{P}_{\mathcal{S}} $ (and $ \mathcal{P}_{\delta P_{nad}} $) which relates to the behaviour of the remaining canonical scalar field. $ \mathcal{S} $ is seen rising and falling during the remaining e-folds of inflation in Figure~\ref{standpow} and Figure~\ref{kinetphipow}. Instead we see in Figure~\ref{kinetchipow} that $ \mathcal{S} $ continues to decrease until slow-roll is no longer satisfied. The final amplitudes for the entropy and non-adiabatic pressure perturbations are significantly larger than for the two standard scalar fields case. The values for the entropy and non-adiabatic pressure perturbation amplitudes are the same as those found in the previous case (where the $\chi$-field reaches the potential's minimum before the $\phi$-field).

There appears to be no difference in the curvature and entropy amplitudes between the two cases considered here in Section~\ref{sec:kinetic}.

\subsubsection{A Scalar Field and DBI Field}
As with the previous model containing the kinetic coupling, this DBI model will also be associated with the same two scenarios, one in which the DBI field decays before the scalar field and the reverse. For all the cases considered, the parameters that belong to the DBI model hold the following values, $ \lambda = 2.0 \times 10^{12} $ and $ \mu = 0.2 $ \cite{SilvTong04}.
 
First we consider the case where inflation is initially dominated by the scalar field $ \phi $. In this scenario, $ \Gamma = 2.0 $ with $ m_{\phi} = 115.6 \times 10^{-7} $ and the initial values for the two fields are $ \chi_0 = 2.0 $ and $ \phi_0 = 12.0 $.
These parameters were chosen so the curvature perturbation amplitude agrees with WMAP measurements.
\begin{figure}[H]
\begin{center}
\scalebox{1.27}{\includegraphics{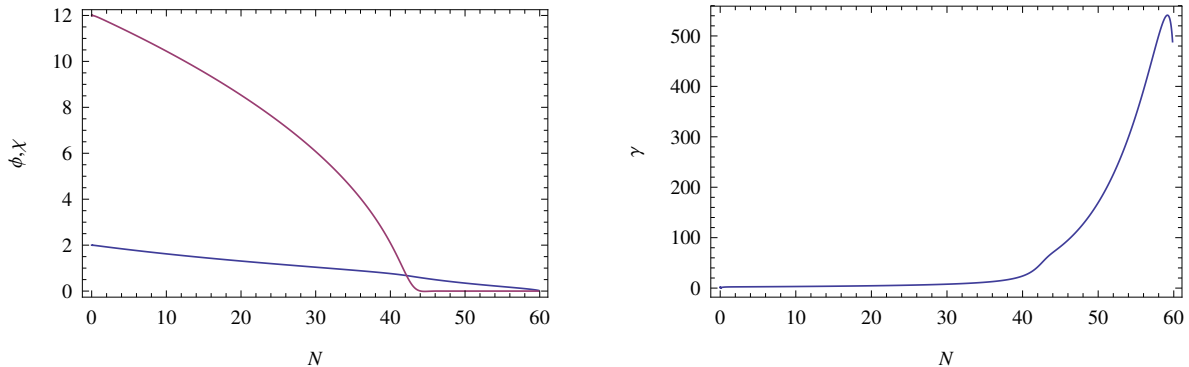}}
\caption{The background dynamics of both the $\phi$-field (purple) and $\chi$-field (blue) are shown on the left. In this scenario, the DBI field is the first to fall to the minimum of the potential well. On the right shows the evolution of the parameter $ \gamma $ during inflation. }
\label{dbiphibggam}
\end{center}
\end{figure}

\begin{figure}[H]
\begin{center}
\scalebox{1.27}{\includegraphics{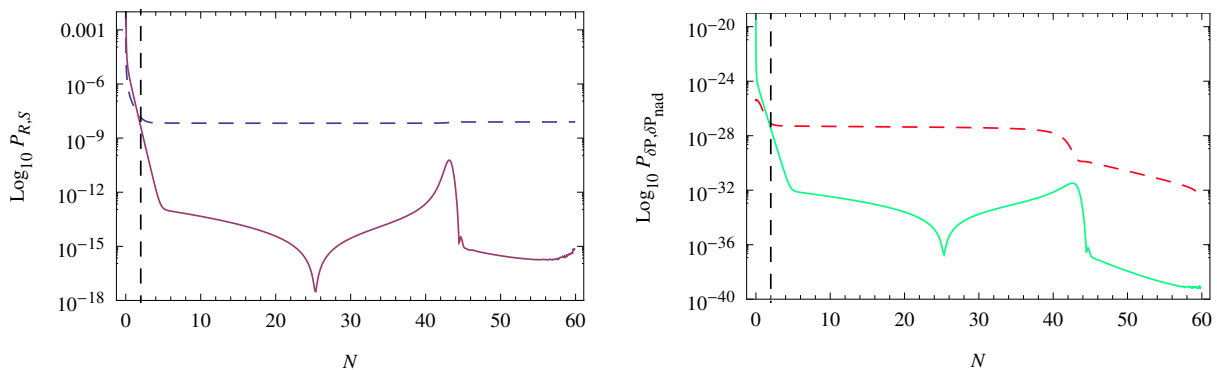}}
\caption{A comparison of the power spectra; on the left are $ \mathcal{P}_{\mathcal{R}} $ (blue, dashed) and $ \mathcal{P}_{\mathcal{S}} $ (purple) and the right, $ \mathcal{P}_{\delta P} $ (red, dashed) and $ \mathcal{P}_{\delta P_{nad}} $ (green). }
\label{dbiphipow}
\end{center}
\end{figure}

From Figure~\ref{dbiphibggam}, there is an exchange in the field contributions at $ N = 44 $ and this is displayed in Figure~\ref{dbiphipow} through the fall in the amplitude of the entropy perturbation. Unlike all the previous actions that were examined, the entropy and non-adiabatic pressure perturbation amplitudes for this particular model of DBI inflation, does not decrease during the final few e-folds of inflation. Furthermore, the final value of the entropy amplitude, $ \mathcal{S} \sim 10^{-16} $ is markedly greater than found in Section~\ref{sec:twostandard}. Similarly, this increase in amplitude is also found in the non-adiabatic pressure perturbation $ \mathcal{P}_{\delta P_{nad}} \sim 10^{-40} $, whereas for the two scalar field case $ \mathcal{P}_{ \delta P_{nad}} \sim 10^{-54} $.

In the other case where the DBI field decays before the scalar field, parameter values are $ \Gamma = 35.1 $ where the mass of the $ \phi$-field is $65 \times 10^{-8} $. The fields have the same starting values as in the previous DBI field case.
\begin{figure}[H]
\begin{center}
\scalebox{1.27}{\includegraphics{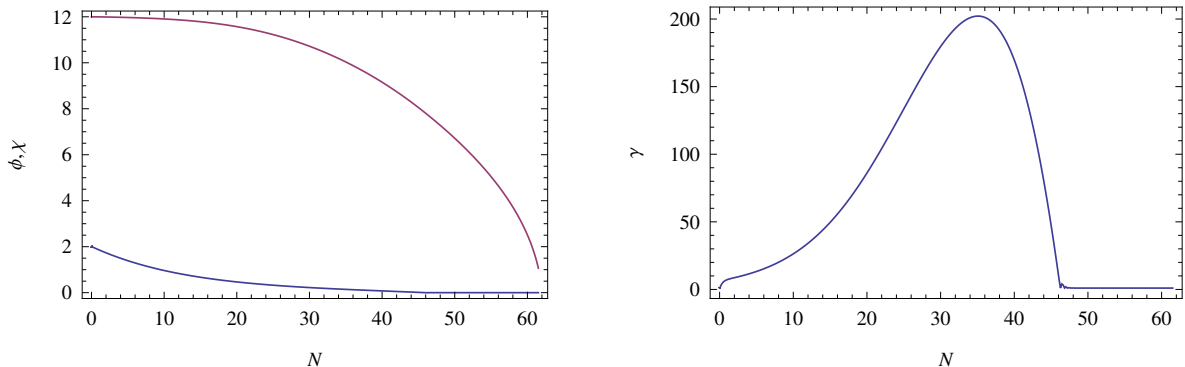}}
\caption{The background dynamics of both the $\phi$-field (purple) and $\chi$-field (blue) are displayed on the left. In this case, the DBI field is the first to fall to the minimum of the potential well. The evolution of $ \gamma $ is shown on the right. }
\label{dbichibggam}
\end{center}
\end{figure}
At first, the DBI field dominates the inflationary period until it reaches the minimum of the potential well and oscillates, at which the $\phi$-field will become dominant.  This is shown in Figure~\ref{dbichipow} through the drop in the amplitude of the entropy perturbation 12 e-folds before the end of inflation.

\begin{figure}[H]
\begin{center}
\scalebox{1.27}{\includegraphics{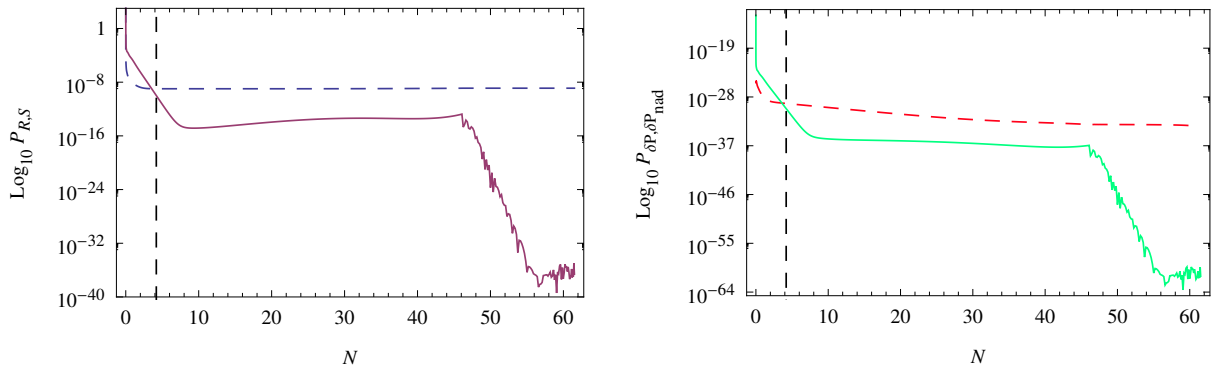}}
\caption{A comparison of the power spectra. $ \mathcal{P}_{\mathcal{R}} $ (blue, dashed) and $ \mathcal{P}_{\mathcal{S}} $ (purple) are shown on the left and $ \mathcal{P}_{\delta P} $ (red, dashed) and $ \mathcal{P}_{\delta P_{nad}} $ (green) are on the right. }
\label{dbichipow}
\end{center}
\end{figure}

The behaviour of all power spectra are similar to those seen in the two canonical fields and kinetic coupling models, where the $ \chi $-field is the first to reach the potential's minimum. This is due to the remaining $\phi$-field experiencing slow-roll. As expected, the drop in the entropy power spectrum at $ N = 47 $ is due to the DBI field reaching the potential's minimum. Final amplitudes for the entropy and pressure (non-adiabatic) perturbations are $ \mathcal{P}_{\mathcal{S}} \sim 10^{-37} $ and $ \mathcal{P}_{\delta P_{nad}} \sim 10^{-62} $. 

Both these values differ greatly than those found for the previous case (when the $\phi$-field is the first to fall to the minimum of the potential). Furthermore, the final amplitudes for $ \mathcal{S} $ and $ \delta P_{nad} $ for this scenario are much smaller than in the two scalar field case in Section~\ref{sec:twostandard}.

\section{Conclusion}
\label{sec:conclusion}
In this paper we studied the evolution of the non-adiabatic pressure perturbation $\delta P_{\rm nad}$ in multi--field inflation. We considered three models containing two massive scalar fields, but each model differs by the form of the kinetic terms. In all models, inflation is driven initially by the two fields. However, one of the field approaches the minimum of its potential before the end of inflation, so that inflation subsequently is driven purely by the second field. Our numerical results for the model with standard kinetic terms agree with \cite{Hust12}. Our main results can be summarised as follows:

\begin{itemize}
\item The nature of kinetic terms affect the evolution and the final value of  $\delta P_{\rm nad}$. In the case of the fields coupled by kinetic terms (model 2 with $\beta$ non--zero) the amplitude of  $\delta P_{\rm nad}$ does not depend on which of the field approaches zero first. However, the amplitude is roughly five orders of magnitude larger for $\beta=0.1$ when compared to the canonical case ($\beta=0$).

\item The evolution of $\mathcal{P}_{\cal S}$ is affected by how fast the first field approaches the minimum of its potential. We generically observe an increase in $\mathcal{P}_{\cal S}$, which influences the results of $\mathcal{P}_{\mathcal{R}}$. We find this effect to be largest in the theory with standard--kinetic terms. In all other models, the first field approaches zero much more gradually, resulting in only a slight increase in $\mathcal{P}_{\cal S}$.  

\item In the third model, if the DBI field drives the last $e$--folds of inflation, we find that $\delta P_{\rm nad}$ is substantially larger than in other cases. On the other hand, if the DBI field is not significant in the later stages of inflation the results are comparable to the standard case, as expected. In this model, the amplitude of $\mathcal{P}_{\cal S}$ is largest.
\end{itemize}

The relevance of our results come from the fact that  $\delta P_{\rm nad}$ sources vorticity at second order, which affects predictions for the $B$--mode polarisation of the CMB. The model with the DBI field driving the last $e$--folds of inflation predicts the largest amplitude for $\delta P_{\rm nad}$ at the end of inflation, which implies that this model predicts a larger source for vorticity than the other models. A comprehensive analysis of pre-- and reheating in these models is necessary to estimate the amount of entropy perturbations in the radiation dominated epoch. The results will be model--dependent, because the details depend on the coupling of the inflaton--field(s) to the matter fields.

\acknowledgements The work of CvdB is supported by the Lancaster-Manchester-Sheffield Consortium for Fundamental Physics under STFC grant ST/J000418/1. SV is supported by a STFC doctoral fellowship. 

\appendix

\section{One Standard Scalar Field and One Field containing a Kinetic Coupling}
\subsection{Perturbation Equations}
\noindent The perturbed Einstein field equations for this model are
\begin{align}
\dot{\Psi} + H \Psi = {} & \frac{1}{2} ( \dot{\phi} \delta \phi + e^{2b} \dot{\chi} \delta \chi ) , \\
\ddot{\Psi} + 4H\dot{\Psi} + ( 2\dot{H} + 3H^2 )\Psi = {} &\frac{1}{2} \bigg[ \dot{\phi} \delta \dot{\phi} + e^{2b} \dot{\chi} \delta \dot{\chi} + b_{\phi} e^{2b} \dot{\chi}^2 \delta \phi - V_{\phi} \delta \phi - V_{\chi} \delta \chi - \dot{\phi}^2 \Psi - e^{2b} \dot{\chi}^2 \Psi \bigg] , \\
3H( H\Psi + \dot{\Psi} ) + \frac{k^2}{a^2} \Psi = {} & - \frac{1}{2} \bigg[ \dot{\phi} \delta \dot{\phi} + e^{2b} \dot{\chi} \delta \dot{\chi} + b_{\phi} e^{2b} \dot{\chi}^2 \delta \phi + V_{\phi} \delta \phi + V_{\chi} \delta \chi -\dot{\phi}^2 \Psi - e^{2b}\dot{\chi}^2 \Psi \bigg] .
\end{align}

\noindent The perturbation equations for the two fields are 
\begin{align}
\label{perturbationkinetic1}
\ddot{Q}_{\phi} {} & + 3H\dot{Q}_{\phi} - 2 b_{\phi} e^{2b} \dot{\chi} \dot{Q}_{\chi} + \bigg(\frac{k^{2}}{a^2} + C_{\phi \phi} \bigg)Q_{\phi} + C_{\phi \chi}Q_{\chi} = 0 ,\\
\label{perturbationkinetic2}
\ddot{Q}_{\chi} {} & + \bigg( 3H + 2b_{\phi} \dot{\phi} \bigg) \dot{Q}_{\chi} + 2 b_{\phi} \dot{\chi} \dot{Q}_{\phi} + \bigg{(}\frac{k^{2}}{a^2} + C_{\chi \chi}\bigg{)}Q_{\chi} + C_{\chi \phi}Q_{\phi} = 0 .
\end{align}
and the coefficients in the equations $ C_{\phi \phi}, C_{\phi \chi}, C_{\chi \chi}, C_{\chi \phi} $ are given as
\begin{align}
C_{\phi \phi} = {} & 3\dot{\phi}^{2} - 2 b_{\phi}^2 e^{2b} \dot{\chi}^2 - \frac{e^{2b} \dot{\phi}^2 \dot{\chi}^2}{2H^2} - \frac{\dot{\phi}^{4}}{2H^{2}}  - b_{\phi \phi} e^{2b} \dot{\chi}^2 
			  + 2\frac{\dot{\phi}}{H}V_{\phi} + V_{\phi \phi} \\
C_{\phi \chi} = {} & 3 e^{2b} \dot{\phi}\dot{\chi} - \frac{ e^{4b} \dot{\phi}\dot{\chi}^{3} }{2H^{2}} - \frac{ e^{2b} \dot{\phi}^{3}\dot{\chi}}{2H^{2}} + \frac{\dot{\phi}}{H}V,_{\chi} + \frac{ e^{2b}\dot{\chi}}{H}V_{\phi} + V_{\phi \chi} \\
C_{\chi \chi} = {} & 3 e^{2b} \dot{\chi}^2 - \frac{ e^{2b} \dot{\chi}^{4} }{2H^{2}} - \frac{ e^{2b} \dot{\phi}^{2}\dot{\chi}^{2} }{2H^{2}} + \frac{2 \dot{\chi}}{H}V_{\chi} + e^{-2b} V_{\chi \chi} \\
C_{\chi \phi} = {} & 3 \dot{\phi} \dot{\chi} - \frac{e^{2b} \dot{\phi} \dot{\chi}^3}{2 H^2} - \frac{ \dot{\phi}^3 \dot{\chi} }{2 H^2 } + 2 b_{\phi \phi} \dot{\phi} \dot{\chi} - 2 b_{\phi}e^{-2b} V_{\chi} 
			  + \frac{e^{-2b} \dot{\phi}}{H} V_{\chi} + \frac{ \dot{\chi} }{H} V_{\phi} + e^{-2b} V_{\phi \chi}
\end{align}

All these equations will reduce to those for the two canonical scalar fields when the kinetic coupling $ b(\phi) $ is set to zero, in this case $ \beta = 0 $.

\addcontentsline{toc}{section}{References}
\bibliography{refs}
\bibliographystyle{ieeetr}

\end{document}